\documentstyle[12pt,aaspp4]{article}
\lefthead{JENKINS ET AL.}
\righthead{IONIZATION OF THE LOCAL INTERSTELLAR MEDIUM}
\begin{document}
\title{The Ionization of the Local Interstellar Medium, as Revealed by
FUSE Observations of N, O and Ar toward White Dwarf
Stars\altaffilmark{1}}
\altaffiltext{1}{Based on data obtained for the Guaranteed Time Team by
the NASA-CNES-CSA FUSE mission operated by the Johns Hopkins University. 
Financial support to U.S. participants has been provided by NASA
contract NAS5-32985.}
\author{E.~B.~Jenkins\altaffilmark{2},
W.~R.~Oegerle\altaffilmark{3},
C.~Gry\altaffilmark{4,5},
J.~Vallerga\altaffilmark{6},
K.~R.~Sembach\altaffilmark{3},
R.~L.~Shelton\altaffilmark{3},
R.~Ferlet\altaffilmark{7},
A.~Vidal-Madjar\altaffilmark{7},
D.~G.~York\altaffilmark{8},
J.~L.~Linsky\altaffilmark{9},
K.~C.~Roth\altaffilmark{3},
A.~K.~Dupree\altaffilmark{10}
and J.~Edelstein\altaffilmark{6}}
\altaffiltext{2}{Princeton University Observatory, Princeton, NJ
08544-1001}
\altaffiltext{3}{Dept.\ of Physics and Astronomy, Johns Hopkins
University, 3400 N. Charles St. Baltimore, MD 21218-2686}
\altaffiltext{4}{ISO Data Center, ESA Astrophysics Division, PO Box
50727, 28080 Madrid, Spain}
\altaffiltext{5}{Laboratoire d'Astronomie Spatiale, B.P.8, 13376
Marseille cedex 12, France}
\altaffiltext{6}{Space Sciences Laboratory, University of California,
Berkeley, CA 94720-7450}
\altaffiltext{7}{Institut d'Astrophysique de Paris, 92 bis, Blvd. Arago,
Paris 7504, France}
\altaffiltext{8}{Dept. of Astronomy and Astrophysics, University of
Chicago, 5640 Ellis Ave., Chicago, IL 60637}
\altaffiltext{9}{JILA, Univ. of Colorado and NIST, Boulder, CO
80309-0440}
\altaffiltext{10}{Smithsonian Astrophysical Observatory, 60 Garden St.,
Cambridge, MA 02138}
\begin{abstract}
FUSE spectra of the white dwarf stars G191-B2B, GD~394, WD~$2211-495$
and WD~$2331-475$ cover the absorption features out of the ground
electronic states of N~I, N~II, N~III, O~I and Ar~I in the far
ultraviolet, providing new insights on the origin of the partial
ionization of the Local Interstellar Medium (LISM), and for the case of
G191-B2B, the interstellar cloud that immediately surrounds the solar
system.  Toward these targets the interstellar abundances of Ar~I, and
sometimes N~I, are significantly below their cosmic abundances relative
to H~I.  In the diffuse interstellar medium, these elements are not
likely to be depleted onto dust grains.  Generally, we expect that Ar
should be more strongly ionized than H (and also O and N whose
ionizations are coupled to that of H via charge exchange reactions)
because the cross section for the photoionization of Ar~I is very high. 
Our finding that Ar~I/H~I is low may help to explain the surprisingly
high ionization of He in the LISM found by other investigators.  Our
result favors the interpretation that the ionization of the local medium
is maintained by a strong EUV flux from nearby stars and hot gases,
rather than an incomplete recovery from a past, more highly ionized
condition.
\end{abstract}
\section{Introduction}\label{intro}

Our solar system is traveling through a low-density ($n_H\sim
0.1-0.3\,{\rm cm}^{-3}$), warm ($T\approx 7000$K), partially ionized gas
cloud called the Local Interstellar Cloud (LIC)  (Lallement et al. 1996)
with a diameter of about 4~pc  (Redfield \& Linsky 2000). Near the LIC
are similar clouds that give rise to absorption features with different
velocities  (Gry 1996; Lallement 1996), known collectively as the Local
Interstellar Medium (LISM).  From the available evidence, the physical
conditions in the LISM clouds are only sightly different from those of
the LIC  (Linsky 1996; Ferlet 1999).

Presently, we do not have a clear understanding of the physical
processes that are responsible for high fractional ionization of helium
in the LISM.  From observations that $\langle n({\rm He~I})/n({\rm
H~I})\rangle=0.07$  (Dupuis et al. 1995), along with the knowledge that
${\rm He/H}=0.10$ in all forms, it is evident that He is usually
slightly more ionized than H.  Also, helium shows less variability in
how strongly it is ionized from one region to the next  (Wolff, Koester,
\& Lallement 1999). While the EUV radiation from white dwarf and other
stars can explain the observed fractional ionization of H  (Vallerga
1998), these sources do not produce enough photons with energies above
24.6~eV to explain the ionization of He.  To resolve this problem, two
main proposals have been offered.  One is that the LISM is not in a
steady-state condition; it is returning from a much more highly ionized
state produced by an energetic event in the recent past ($t\lesssim
10^6\,{\rm yr}$), such as the flash from a nearby supernova  (Reynolds
1986; Frisch \& Slavin 1996) or its shock wave  (Lyu \& Bruhweiler
1996).  The other possibility is that the ionization of He is maintained
in a steady state by the diffuse EUV radiation arising from conductive
interfaces between the cloud edges and the surrounding, hot medium at
$T\sim 10^6$K  (Slavin 1989; Slavin \& Frisch 1998), or, alternatively,
by recombination radiation from highly ionized but cooled gases that
might surround us  (Breitschwerdt \& Schmutzler 1994).

Sofia \& Jenkins  (1998) have proposed that the apparent abundances of
the neutral forms of N, O and Ar, elements that are generally very
lightly depleted in the interstellar medium (if at all), can help to
unravel the mystery of the He ionization.  UV absorption lines in the
spectra of white dwarf stars present good opportunities for studying
these abundances in the LISM, since most of the brighter objects are
well removed from major gas clouds elsewhere and remain immersed within
a ``Local Bubble'' (diam. $\sim 200$~pc) of very hot, low density
material surrounding the nearby cloud complex  (Cox \& Reynolds 1987;
Sfeir et al. 1999).  We report here on initial spectra of four white
dwarf stars taken with the Far Ultraviolet Spectroscopic Explorer (FUSE)
(Moos et al. 2000; Sahnow et al. 2000) taken at a resolution of
$20-25\,{\rm km~s}^{-1}$ (\S\S\ref{obs}$-$\ref{results}), and we
interpret the findings in the light of the Sofia \& Jenkins proposal to
obtain a better understanding on how the LISM is ionized
(\S\ref{interpretation}).

\section{Observations}\label{obs}

The stars, their locations in the sky, estimates for $N$(H~I) arising
from the LIC only, and the wavelength ranges observed by FUSE are
summarized in Table~\ref{stars}.  All of the stars are at distances that
range between approximately 50$-$80~pc.  The values of $N({\rm
H~I})_{\rm LIC}/N({\rm H~I})$ (cf. Tables~\ref{stars} and
\ref{col_dens}) indicate that the line of sight to G191-B2B contains
mostly material from the LIC, while those toward the remaining three
stars are dominated by other clouds in the LISM.

\placetable{stars}
\begin{deluxetable}{
c     
c     
c     
c     
c     
}
\tablecolumns{5}
\tablewidth{0pt}
\tablecaption{Target Stars, Locations and Wavelengths
Ranges\label{stars}}
\tablehead{
\colhead{Star\tablenotemark{a}} & \colhead{$\ell$} & \colhead{$b$} &
\colhead{$\log N({\rm H~I})$}
& \colhead{$\lambda$ Coverage}\\
\colhead{} & \colhead{(deg)} & \colhead{(deg)} & \colhead{for the
LIC\tablenotemark{b}}
& \colhead{(\AA)}
}
\startdata
G191-B2B&156.0&+7.1&$18.25$&$987-1082$\nl
GD 394&91.4&1.1&17.68&$987-1082$\nl
WD $2211-495$&345.8&$-52.6$&16.73&$905-1082$\nl
WD $2331-475$&334.9&$-64.8$&17.12&$905-1082$\nl
\enddata
\tablenotetext{a}{Ordered according to increasing foreground $N$(H~I) --
see Table~\protect\ref{col_dens}}
\tablenotetext{b}{Estimate for the contribution to the hydrogen column
that arises from the Local Interstellar Cloud surrounding our solar
system, according to Redfield \& Linsky  (2000).}
\end{deluxetable}

\begin{figure}
\plotone{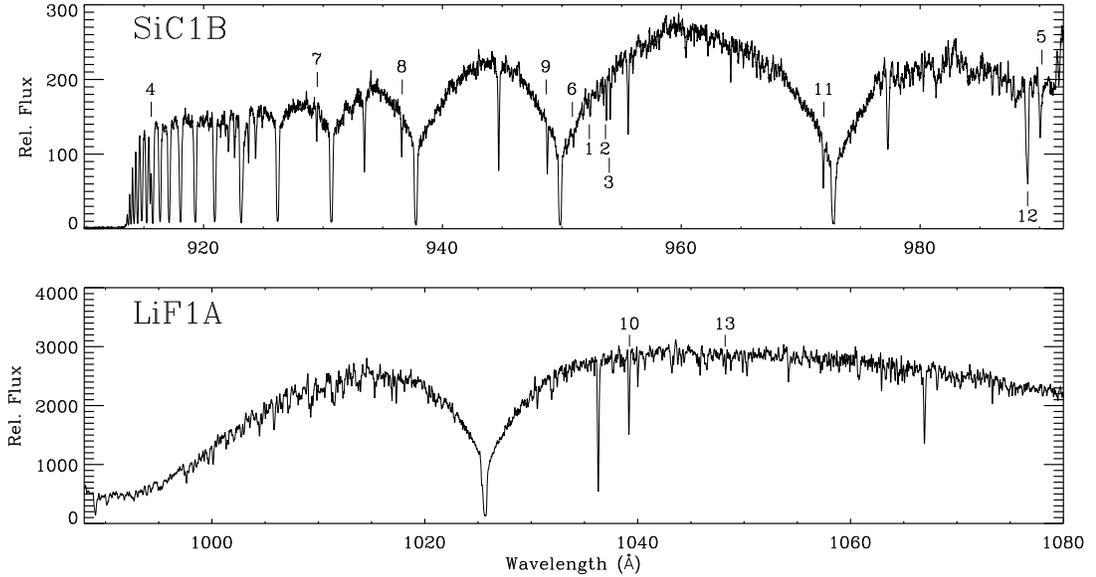}
\caption{Far ultraviolet spectra of WD~2211$-$495 recorded by FUSE.
Important features for our study are identified according to the numbers
listed in the last column of Table~\protect\ref{meas}.\label{wd2211}}
\end{figure}

Our spectra cover the important absorption features from N, O and Ar in
their neutral forms.  In addition, for some stars we recorded two
ionized forms of nitrogen, N~II and N~III.  While the N~III feature
conceivably might arise from photospheric absorption, we see no evidence
of stronger absorption from N~III in an excited fine-structure level --
in the absence of saturation this feature would be twice as strong as
the one we observed if all of the N~III absorption came from the star's
photosphere.  Two spectral segments for one of the stars, WD~$2211-495$,
are shown in Fig.~\ref{wd2211}.

\placefigure{wd2211}

\begin{deluxetable}{
c     
c     
c     
c     
c     
c     
c     
c     
}
\small
\tablecolumns{6}
\tablewidth{525pt}
\tablecaption{FUSE Measurements of Absorption Lines\label{meas}}
\tablehead{
\colhead{Species} & \colhead{$\lambda$} &
\colhead{$\log (f\lambda)$\tablenotemark{a}} &
\colhead{WD~$2211-495$} &
\colhead{WD~$2331-475$} &
\colhead{Ident.}\\
\colhead{} & \colhead{(\AA)} & \colhead{}
& \colhead{[$W_\lambda$(m\AA)]}
& \colhead{[$W_\lambda$(m\AA)]}
& \colhead{in Fig.~\protect\ref{wd2211}}
}
\startdata
N~I \dotfill&952.303&0.532\tablenotemark{b}&$16.9\pm2.4$&
\nodata&1\nl
&965.041&0.589&\nodata&$14.6\pm4.2$&\nodata\nl
&954.104&0.809&$4.8\pm1.5$&\nodata&\nodata\nl
&964.626&0.959&\nodata&$23.4\pm4.1$&\nodata\nl
&953.415&1.098&\nodata&$34.6\pm4.5$&\nodata\nl
&963.990&1.154&\nodata&$22.4\pm4.2$&\nodata\nl
&953.655&1.376&$24.5\pm2.1$&$34.4\pm4.1$&2\nl
&953.970&1.521&$23.2\pm2.0$&$47.4\pm4.8$&3\nl
N~II\dotfill&915.612&2.180&$89.\pm15.$\tablenotemark{c}&
$59.\pm20.$\tablenotemark{c}&4\nl
N~II$^*$\dotfill&916.012&2.180\tablenotemark{b}
&$-2.7\pm2.9$\tablenotemark{d}&
$-2.9\pm5.3$\tablenotemark{d}&\nodata\nl
N~II$^{**}$\dotfill&916.701&2.182\tablenotemark{b}
&$3.6\pm2.8$\tablenotemark{d}&
$-9.0\pm5.7$\tablenotemark{d}&\nodata\nl
N~III\dotfill&989.799\tablenotemark{e}&2.085&
$64.5\pm4.0$&$70.9\pm4.0$&5\nl
N~III$^*$\dotfill&991.577\tablenotemark{f}&2.085\tablenotemark{b}
&$26.1\pm2.3$&$7.8\pm4.4$&\nodata\nl
O~I\dotfill&950.884&0.176&$13.4\pm2.3$&
$29.1\pm5.6$&6\nl
&929.517&0.329&$27.3\pm3.0$&\nodata&7\nl
&1026.476\tablenotemark{g}&0.434&$4.3\pm2.1$&
$-0.01\pm3.0$&\nodata\nl
&976.448&0.509&$7.5\pm2.3$&$58.2\pm6.0$&\nodata\nl
&936.630&0.534&$32.5\pm2.4$&$40.6\pm4.7$&8\nl
&948.685&0.778&$50.4\pm3.0$&$69.9\pm6.9$&9\nl
&1039.230\tablenotemark{h}&0.980&$67.5\pm1.5$&
$93.8\pm1.8$&10\nl
&971.738&1.128&$55.5\pm3.6$&$124.0\pm13.2$&11\nl
&988.773\tablenotemark{i}&1.737\tablenotemark{b}&
$162.5\pm3.0$&$184.6\pm7.5$&12\tablebreak
Ar~I\tablenotemark{j}\dotfill&1048.218\tablenotemark{k}&2.440&$15.1\pm1.0$&
$22.5\pm1.8$&13\nl
Ar~II\dotfill&919.781&0.911&\nodata&$0.3\pm6.1$&\nodata\nl
\enddata
\tablenotetext{a}{From a privately distributed update by D.~C.~Morton to
his compilation published earlier  (Morton 1991).}
\tablenotetext{b}{Two or more lines blended; line strength is for all of
them combined.}
\tablenotetext{c}{The large uncertainty is caused by an uncertain
continuum in the vicinity of the closely spaced higher Lyman series
transitions of H~I.}
\tablenotetext{d}{In principle, our upper limits for the column
densities of N~II in the excited fine structure levels, N~II$^*$ and
N~II$^{**}$, may be compared to lower limits for the amount of unexcited
N~II to arrive at a limit for the representative electron density 
(McKenna et al. 1996).  Toward WD~$2211-495$ and WD~$2331-475$ we
conclude that if $T=7000$K, $n(e)< 0.8$ and $2.0\,{\rm cm}^{-3}$,
respectively, but these values are higher than our actual expectations
for the LISM based on C~II$^*$/C~II and Mg~I/Mg~II seen elsewhere 
(Lallement \& Ferlet 1997; Wood \& Linsky 1997; Jenkins, Gry, \& Dupin
2000).}
\tablenotetext{e}{G191-B2B: $51.0\pm 4.1$m\AA; GD~394: $15.5\pm
10.9$m\AA.  This line could have interference from an absorption by
Si~II $\lambda 989.873$, but it is not likely to be strong.}
\tablenotetext{f}{G191-B2B: $-2.4\pm 4.0$m\AA; GD~394: $15.4\pm
10.0$m\AA.}
\tablenotetext{g}{G191-B2B: $1.3\pm 2.7$m\AA; not measured for GD~394. 
This line consistently gave lower than expected absorption, based on the
outcomes from other lines and the relative $f$-values.  Hence, the
published $f$-value for this line may be too large.  For this reason, we
disregarded indications from this line.}
\tablenotetext{h}{G191-B2B: $35.5\pm 1.8$m\AA; GD~394: $51.9\pm
3.4$m\AA.}
\tablenotetext{i}{G191-B2B: $88.4\pm 3.9$m\AA; GD~394: $139.4\pm
11.0$m\AA.}
\tablenotetext{j}{The line of Ar~I at 1066.66\AA\ was not measured
because of possible confusion with photospheric features of Si~IV at
1066.61, 1066.64 and 1066.65\AA.}
\tablenotetext{k}{G191-B2B: $4.1\pm 1.5$m\AA; GD~394: $12.0\pm
2.9$m\AA.}
\end{deluxetable}
\clearpage

\section{Results}\label{results}

The first and second columns in Table~\ref{meas} show the species and
the wavelengths of their transitions that we measured within the broad
spectral coverages for WD~$2211-495$ and WD~$2331-475$, with the lines
ranked according to their strengths expressed in terms of
$\log(f\lambda)$ (column 3) in each group.  The fourth and fifth columns
show the equivalent widths (in m\AA) and their errors arising from noise
and uncertainties in the continuum and background levels.  Our noise
estimate was derived from the sizes of flux deviations from the adopted
continuum level and thus included the effect of counting statistical
fluctuations and small variations in the response of the detector
(flat-field corrections were not made).  Results for the more limited
set of lines that we could measure for G191-B2B and GD~394 are given in
the endnotes of the table.

\placetable{meas}

White dwarf stars can have narrow, photospheric features that could
interfere with or masquerade as interstellar lines.  This problem is
probably more severe for the stars GD~394 and WD~$2211-495$, which are
known to be metal rich  (Wolff et al. 1998).  For example, the
952.3$\,$\AA\ feature of N~I in WD~$2211-495$ is probably enhanced by a
stellar feature, since the observed equivalent width is far higher than
those of other transitions of comparable strength.

Table~\ref{col_dens} summarizes our best estimates for the column
densities of N~I, O~I, and Ar~I, along with the values for $N$(H~I)
reported by other investigators.  A velocity dispersion $b=7\,{\rm
km~s}^{-1}$ was derived for the O~I lines toward WD~$2211-495$.  The
equivalent widths for the N~I lines for this star are consistent with
this $b$-value.  Toward WD~$2331-475$ the N~I lines give a best fit to
$b=6\,{\rm km~s}^{-1}$, while the O~I lines are more consistent with 
$b=10\,{\rm km~s}^{-1}$.  This behavior probably arises from strong
contrasts in physical environments for two or more velocity
subcomponents, leading to different ratios of O~I to N~I in each case. 
For cases where the lines could not be seen or were marginally detected,
we list upper limits for the column densities computed for the
measurements plus their $2\sigma$ errors, assuming that the lines have
no saturation.  For such cases, a large wavelength interval ($\sim
0.3$\AA) had to be considered (thus increasing the errors), because our
wavelength scale was not very accurate.

\placetable{col_dens}
\begin{deluxetable}{
c     
c     
c     
c     
c     
}
\tablecolumns{5}
\tablewidth{525pt}
\tablecaption{Neutral Atoms: Logarithms of Column Densities (cm$^{-2}$)
and Deficiencies\tablenotemark{a}\label{col_dens}}
\tablehead{
\colhead{Species} & \colhead{G191-B2B} & \colhead{GD 394} &
\colhead{WD~$2211-495$} & \colhead{WD~$2331-475$}\\
}
\startdata
H~I\dotfill&18.36\tablenotemark{b}
&$18.65_{-0.34}^{+0.16}$\tablenotemark{c}
&18.76\tablenotemark{d}&18.93\tablenotemark{d}\nl
N~I\dotfill&13.90\tablenotemark{b}&$13.85\pm0.15$\tablenotemark{e}&
$14.02\pm 0.15$&$14.61\pm0.15$\nl
&($-0.26$)&($-0.60$)&($-0.54$)&($-0.12$)\nl
&\nl
O~I\dotfill&14.84\tablenotemark{b}&$14.94\pm0.20$\tablenotemark{f}&
$15.32^{+0.1}_{-0.3}$&$15.45\pm0.1$\nl
&($-0.19$)&($-0.38$)&($-0.11$)&($-0.15$)\nl
&\nl
Ar~I\dotfill&$<12.44$&
$12.70\pm0.1$&$12.83\pm0.1$&
$13.06\pm0.1$\tablenotemark{g}\nl
&($<-0.42$)&($-0.45$)&($-0.43$)&($-0.37$)\nl
\enddata
\tablenotetext{a}{Values in parentheses indicate by what logarithmic
amount the abundances relative to H~I deviate from their respective
values in B stars, for which N = 7.80  (Cunha \& Lambert 1994), O = 8.67 
(Cunha \& Lambert 1992) and Ar = 6.50  (Holmgren et al. 1990; Keenan et
al. 1990) on a scale where H = 12.00.}
\tablenotetext{b}{ (Vidal-Madjar et al. 1998).}
\tablenotetext{c}{ (Barstow et al. 1996).}
\tablenotetext{d}{ (Wolff et al. 1998); no errors stated.}
\tablenotetext{e}{From an observation of the N~I multiplet at
1200$\,$\AA\ with GHRS on HST, reported by Barstow et al.  (1996). 
While an arbitrarily large value of $b$ gives a best fit to the lines,
we expect that $b\approx 10\,{\rm km~s}^{-1}$ is more reasonable (and
still gives an acceptable fit).}
\tablenotetext{f}{Based on our observations of the O~I lines, but using
$b=10\,{\rm km~s}^{-1}$ (see note $e$ above).}
\tablenotetext{g}{$b=6\,{\rm km~s}^{-1}$ was adopted, as indicated by
N~I.  The result drops to 13.01 if $b=10\,{\rm km~s}^{-1}$ (from O~I).}
\end{deluxetable}

Our ultimate objective was to judge the column densities in the light of
a simple expectation based on some meaningful cosmic abundance scale.
The entries within parentheses in Table~\ref{col_dens} show how strongly
deficient the neutral forms of these elements are with respect to their
expected abundances relative to hydrogen, using the abundances in B
stars as a standard.

\section{Interpretation}\label{interpretation}

The deficiency of Ar~I for all four stars is of order $-0.4$~dex.  We
adopt the argument presented by Sofia \& Jenkins  (1998) that this is
not caused by Ar depleting onto the surfaces of dust grains, but rather
that neutral Ar is more easily photoionized than H.  If some photons
with $E>13.6\,$eV can penetrate a region, the H can be partially
ionized.  Under these circumstances the Ar (I.P.~=~15.76$\,$eV) should
be much more fully ionized because its photoionization cross section is
about ten times that of H over a broad range of energies.

N~I and O~I also have photoionization cross sections that are larger
than that of H~I, but the ionization fractions of N and O are coupled to
that of H via resonant charge exchange reactions.  For O~II, the rate
coefficient for charge exchange with H~I ($\sim 10^{-9}{\rm cm}^3{\rm
s}^{-1}$)  (Field \& Steigman 1971) is much larger than the coefficient
of recombination with free electrons ($4\times 10^{-13}{\rm cm}^3{\rm
s}^{-1}$ for $T=7000$K), so the coupling of the O and H ionization
fractions is very strong, unless there is appreciable ionization of O to
higher stages.  N~II has a charge exchange rate that is only about twice
the recombination coefficient  (Butler \& Dalgarno 1979), so in regions
where $n_{\rm e} \gg n_{\rm H}$, N can start to behave more like Ar and
show a deficiency of its neutral form.

\begin{figure}
\epsscale{0.5}
\plotone{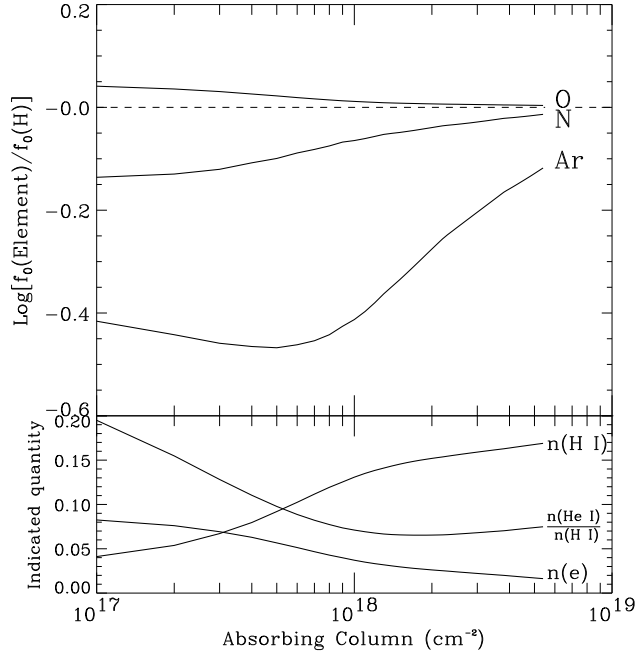}
\caption{Top panel: Logarithms of the expected deficiencies of the
neutral fractions $f_0=n({\rm neutral})/n({\rm total})$ of N, O and Ar,
compared to those of H, caused by photoionization arising from the
radiation by hot stars and a hot-gas conduction front surrounding a LISM
cloud.  These fractions are plotted as a function of the shielding
depths [scaled to $N$(H~I)] within a cloud at a uniform pressure
$p/k=1.5\times 10^3{\rm cm}^{-3}$K and temperature $T=7000$K.  Local
densities of neutral hydrogen and electrons (in ${\rm cm}^{-3}$) and the
ratio of neutral H to He are shown in the bottom panel.\label{calc_def}}
\end{figure}

To illustrate the expected deviations for O~I/H~I, N~I/H~I and Ar~I/H~I
away from the condition where all three elements are fully neutral,
Fig.~\ref{calc_def} shows the results of our calculations of
photoionization equilibria for various depths of shielding by neutral H
and He within a cloud.  The detailed equations and how they are solved
are described by Sofia \& Jenkins  (1998).  At any point within the
cloud the radiation field caused by hot stars was assumed to be stronger
or weaker than Vallerga's  (1998) combined stellar fluxes by a factor
$\exp\{ [9\times 10^{17}{\rm cm}^{-2}-N({\rm H~I})][\sigma(\lambda)_{\rm
H~I} + 0.07\sigma(\lambda)_{\rm He~I}]\}$, where the $\sigma(\lambda)$'s
are the photoionization cross sections.  We supplemented this field with
Slavin's  (1989) calculated flux from a cloud's conductive interface
with much hotter gases.\footnote{We attenuated the strong emission peak
near 70$\,$\AA\ caused by Fe~IX, Fe~X and Fe~XI, recognizing that
McCammon et al.  (2000) failed to observe these lines at a flux level
far below the predicted one.  This probably happens because Fe is
depleted onto dust grains.}

\placefigure{calc_def}

Turning back to the observations, the results shown in
Table~\ref{col_dens} indicate that N and O are deficient, but for
G191-B2B and WD~$2331-475$ not as strongly as Ar.  A mild deficiency in
N can be understood if the hydrogen is mostly ionized, as is expected
near the boundary of a cloud (see Fig.~\ref{calc_def}).  The importance
of ionization in reducing N~I is reinforced by our lower limits
(assuming no saturations of the profiles) for [$\log N({\rm N~II}),~\log
N({\rm N~III})$] = [unavail.,~13.68], [13.86,~13.78], [13.68,~13.82]
toward G191-B2B, WD~$2211-495$ and WD~$2331-475$.  Here, we can state
that {\it at least\/} 38, 56 and 22\% of all the nitrogen is in an
ionized form toward the respective stars.  The underabundances of O are
harder to explain, although they are reduced in magnitude by 0.17~dex if
one adopts as a standard the O/H found in the ISM by Meyer, et al. 
(1998) (8.50 on a logarithmic scale with H~=~12.00) rather than the
value of O/H in B stars.  Also, in view of the possibly large
uncertainties in $N({\rm H~I})$ (except for G191-B2B), it may be better
to adopt O as the reference element instead of H when we investigate the
behavior of N~I and Ar~I.

For GD~394 and WD~$2211-495$, it is difficult to understand why the
deficiencies of N~I are not much weaker than the corresponding ones of
Ar~I.  One possible explanation could be that a conduction front is not
the dominant source of high energy radiation, but instead the ionization
of N and Ar might be strongly influenced by $40-80\,$eV photons arising
from recombinations to He~II, which could conceivably be the dominant
form of diffuse EUV radiation (Breitschwerdt, 2000) if much of the
material in the Local Bubble has cooled to $T\lesssim 10^5$K but has not
yet recombined  (Breitschwerdt \& Schmutzler 1994).  Over much of this
energy range, N~I is more easily ionized than Ar~I if recombinations
with free electrons are more important than charge exchange with H~I
[see Fig.~3 of Sofia \& Jenkins  (1998)].  It will be interesting to see
if detailed calculations confirm this conjecture.

A flux of ionizing photons with E~$>$~24.6~eV and sufficient intensity
to create the steady-state ionization of He in the LISM has not yet been
observed directly, but our observations showing a significant reduction
in Ar~I favor its existence.  If, by contrast, we had found a normal
amount of Ar~I, we might have concluded that the gas was probably more
highly ionized in the past and has not yet fully recombined.  Here, a
near equality of Ar~I/H~I to the expected cosmic ratio would have arisen
from their virtually identical recombination coefficients.  Ionizations
to levels higher than the singly ionized forms are not important in this
picture, since their recombination rates are significantly faster
($\propto Z^2$).

\acknowledgements

We thank T.~M.~Tripp and C.~Howk for their review and comments on an
early draft of this paper.


\begin{references}

\reference{3788} Barstow, M. A., Holberg, J. B., Hubeny, I., Lanz, T., 
Bruhweiler, F. C., \& Tweedy, R. W. 1996, \mnras,  279, 1120

\reference{} Breitschwerdt, D. 2000, private communication

\reference{2877} Breitschwerdt, D., \& Schmutzler, T. 1994, Nat,  371,
774

\reference{1927} Butler, S. E., \& Dalgarno, A. 1979, \apj,  234, 765

\reference{1894} Cox, D. P., \& Reynolds, R. J. 1987, \araa,  25, 303

\reference{3789} Cunha, K., \& Lambert, D. L. 1992, \apj,  399, 586

\reference{3500} --- 1994, \apj,  426, 170

\reference{3171} Dupuis, J., Vennes, S., Bowyer, S., Pradhan, A. K., \& 
Thejll, P. 1995, \apj,  455, 574

\reference{3804} Ferlet, R. 1999, \aapr,  9, 153

\reference{3406} Field, G. B., \& Steigman, G. 1971, \apj,  166, 59

\reference{3356} Frisch, P. C., \& Slavin, J. D. 1996, Space Sci. Rev., 
78,
 223

\reference{3798} Gry, C. 1996, Space Sci. Rev.,  78, 239

\reference{3431} Holmgren, D. E., Brown, P. J. F., Dufton, P. L., \&
Keenan,
 F. P. 1990, \apj,  364, 657

\reference{3724} Jenkins, E. B., Gry, C., \& Dupin, O. 2000, \aap,  354,
253

\reference{1905} Keenan, F. P., Bates, B., Dufton, P. L., Holmgren, D.
E., 
\& Gilheany, S. 1990, \apj,  348, 322

\reference{3797} Lallement, R. 1996, Space Sci. Rev.,  78, 361

\reference{4149} Lallement, R., \& Ferlet, R. 1997, \aap,  324, 1105

\reference{3802} Lallement, R., Linsky, J. L., Lequeux, J., \& Branov,
V. B.
 1996, Space Sci. Rev.,  78, 299

\reference{3799} Linsky, J. L. 1996, Space Sci. Rev.,  78, 157

\reference{3237} Lyu, C. H., \& Bruhweiler, F. C. 1996, \apj,  459, 216

\reference{3791} McCammon, D., Almy, R., Apodaca, E., Bergmann, W.,
Deiker, 
S., Galeazzi, M., Lesser, A., Sanders, W., Figueroa, E., Kelley, R. L., 
Porter, F. S., Stahle, C. K., \& Szymkowiak, A. E. 2000, \baas,   in
press 
(late paper nr. 135.04, AAS Atlanta meeting)

\reference{316} McKenna, F. C., Keenan, F. P., Foster, V. J., Jenkins,
E. 
B., Bell, K. L., \& Stafford, R. P. 1996, in UV and X-ray Spectroscopy
of 
Astrophysical and Laboratory Plasmas, ed. K. Yamashita \& T. Watanabe
(Tokyo: 
Universal Academy Press), p. 499

\reference{4196} Meyer, D. M., Jura, M., \& Cardelli, J. A. 1998, \apj,  
493, 222

\reference{} Moos, H. W. et al. 2000, \apj, this issue

\reference{96} Morton, D. C. 1991, \apjs,  77, 119

\reference{3796} Redfield, S., \& Linsky, J. L. 2000, \apj,   submitted

\reference{2603} Reynolds, R. J. 1986, \aj,  92, 653

\reference{} Sahnow, D. J. et al. 2000, \apj, this issue

\reference{3701} Sfeir, D. M., Lallement, R., Crifo, F., \& Welsh, B. Y. 
1999, \aap,  346, 785

\reference{1773} Slavin, J. D. 1989, \apj,  346, 718

\reference{3790} Slavin, J. D., \& Frisch, P. C. 1998, in The Local
Bubble 
and Beyond, ed. D. Breitschwerdt, M. J. Freyberg, \& J. Tr\"umper
(Berlin: 
Springer), p. 305

\reference{4310} Sofia, U. J., \& Jenkins, E. B. 1998, \apj,  499, 951

\reference{3481} Vallerga, J. 1998, \apj,  497, 921

\reference{3562} Vidal-Madjar, A., Lemoine, M., Ferlet, R., H\'ebrard,
G., 
Koester, D., Audouze, J., Cass\'e, M., Vangioni-Flam, E., \& Webb, J.
1998, 
\aap,  338, 694

\reference{3787} Wolff, B., Koester, D., Dreizler, S., \& Haas, S. 1998, 
\aap,  329, 1045

\reference{3717} Wolff, B., Koester, D., \& Lallement, R. 1999, \aap, 
346, 
969

\reference{4258} Wood, B. E., \& Linsky, J. L. 1997, \apjl,  474, L39

\end{references}
\end{document}